\documentclass[final,3p,times]{elsarticle}

\usepackage[english]{babel}
\usepackage{textcomp}
\usepackage{graphicx}
\usepackage{flafter}
\usepackage{amsmath,amssymb}
\usepackage{wasysym}
\usepackage{bm}
\usepackage[T1]{fontenc}
\usepackage[ansinew]{inputenc}
\usepackage{color}

\newcommand{\il}{~}
\newcommand{\ti}[1]{\mbox{\tiny{#1}}}

\def\be{\begin{equation}}
\def\ee{\end{equation}}
\def\bes{\begin{equation*}}
\def\ees{\end{equation*}}
\def\bea{\begin{eqnarray}}
\def\eea{\end{eqnarray}}

\newcommand{\eref}[1]{Eq.(\ref{#1})}

\newcommand{\erefs}[1]{Eqs.(\ref{#1})}

\newcommand{\reff}[1]{(\ref{#1})}

\newcommand{\p}{\partial}
\newcommand{\etab}{\bar{\eta}}
\newcommand{\pt}{\p_t}
\newcommand{\vna}{\vec\nabla}
\newcommand{\lapl}{\nabla^2}

\renewcommand{\vec}[1]{\boldsymbol{#1}}
\renewcommand{\b}{\vec{b}}
\newcommand{\ro}{\rho_0}
\newcommand{\ru}{\rho_1}
\newcommand{\po}{P_0}
\newcommand{\pu}{P_1}
\newcommand{\phio}{\Phi_0}
\newcommand{\phiu}{\Phi_1}
\newcommand{\bo}{\vec{B}_0}
\newcommand{\bu}{\vec{B}_1}
\newcommand{\vo}{\vec{v}_0}
\newcommand{\vu}{\vec{v}_1}
\newcommand{\hbo}{\hat{\vec{B}}_0}
\newcommand{\hq}{\hat{\vec{q}}}

\newcommand{\para}{{\scriptscriptstyle{\parallel}}}
\newcommand{\perpt}{{\scriptscriptstyle{\perp}}}
\newcommand{\vup}{v_{1}^\para}
\newcommand{\vuo}{\vec{v}_{1}^\perpt}

\newcommand{\omo}{\omega_{\ti{0}}^2}
\newcommand{\oma}{\omega_{\mbox{\tiny{A}}}^2}
\newcommand{\ome}{\omega_{\eta}}

\newcommand{\ie}{\emph{i.e.}, }
\newcommand{\kv}{\vec{q}}
\newcommand{\rv}{\vec{r}}
\newcommand{\bb}{\bar{b}}
\newcommand{\vb}{\bar{v}}

\journal{Physica D}

\begin{document}

\begin{frontmatter}

\title{Stability of a self-gravitating homogeneous resistive plasma}

\author[Sapienza,QM]{Daniela Pugliese}

\author[Sapienza]{Nakia Carlevaro}

\author[Sapienza,Oxford]{Massimiliano Lattanzi}

\author[ENEA,Sapienza,INFN]{Giovanni Montani\corref{cor1}}
\ead{giovanni.montani@frascati.enea.it}
\cortext[cor1]{Corresponding Author. Contact Address: Dipartimento di Fisica, Universit\`a di Roma ``Sapienza'', Piazzale Aldo Moro 5, 00185 Roma (Italy).
Tel: +39 06 49914356.}

\author[Sapienza]{Riccardo Benini}

\address[Sapienza]{Dipartimento di Fisica, ``Sapienza'' Universit\`a di Roma,\\
P.le Aldo Moro 5, 00185 Roma (Italy).}
\address[QM]{{School of Mathematical Sciences, Queen Mary University of London.}}
\address[Oxford]{Oxford Astrophysics, Denys Wilkinson Building, Keble Road OX1 3RH, Oxford (UK)}
\address[ENEA]{ENEA - C.R. Frascati (Rome), UTFUS-MAG.}
\address[INFN]{INFN - Sez. Roma1.}

\begin{abstract}

In this paper, we analyze the stability of a homogeneous self-gravitating plasma, having a non-zero resistivity. This study provides a generalization of the Jeans paradigm for determining the critical scale above which gravitational collapse is allowed.

We start by discussing the stability of an ideal self-gravitating plasma embedded in a constant magnetic field. We outline the existence of an anisotropic feature of the gravitational collapse. In fact, while in the plane orthogonal to the magnetic field the Jeans length is enhanced by the contribution of the magnetic pressure, outside this plane perturbations are governed by the usual Jeans criterium. The anisotropic collapse of a density contrast is sketched in details, suggesting that the linear evolution provides anisotropic initial conditions for the non-linear stage, where this effect could be strongly enforced.

The same problem is then faced in the presence of non-zero resistivity and the conditions for the gravitational collapse are correspondingly extended. The relevant feature emerging in this resistive scenario is the cancellation of the collapse anisotropy in weakly conducting plasmas. In this case, the instability of a self-gravitating resistive plasma is characterized by the standard isotropic Jeans length in any directions. The limit of very small resistivity coefficient is finally addressed, elucidating how reminiscence of the collapse anisotropy can be found in the different value of the perturbation frequency inside and outside the plane orthogonal to the magnetic field.
\end{abstract}


\end{frontmatter}

\section{Introduction}\label{xanes}

In many astrophysical and cosmological systems the presence of a plasma component has a very important role in determining the shape and the behavior of  their equilibrium configurations. The peculiarity of these plasma configurations with respect to those ones observed in laboratory, relies on the dominant character of the gravitational interaction in determining the stability properties. Indeed, as firstly suggested by Jeans \cite{H1,H2}, the gravitational interaction is able to induce the collapse, as long as a critical scale of the configuration is reached. Such a scale, depending on the sound speed and on the mass density of the medium is commonly known as  the \emph{Jeans length}. On the other hand, dealing with a self-gravitating plasma instead of a fluid brings in the equilibrium all the typical features observed in magnetically confined and highly ionized gases, like the emergence of Alfv\'en and magnetosonic waves. Moreover, we stress that the possibility to postulate the presence of a magnetic field, is allowed since it is observationally demonstrated by the direct observation of astrophysical systems
(see \emph{e.g.} Refs. \cite{1988A&A...190...41K,1990ApJ...362..449O,1991ApJ...379...80K,1992ApJ...387..528K,1992ApJ...388...17W,1993ApJ...416..554T,1994ApJ...435L.109L,1995A&A...302..680F} and, more recently, Refs. \cite{2002ApJ...567..202E,2005AJ....130.2566S,2008A&A...478..435R,2009IAUS..256..178M,2009ApJ...705L.176S,2010ApJ...714.1170M,2010JETPL..90..637V,2011IAUS..274..325B}; see also Refs. \cite{Kronberg:1993vk,1997Natur.385..131Z,1996ARA&A..34..155B,Carilli:2001hj,2011NewAR..55...91V} for review works).

The fact that, in our analysis, both the background mass density and magnetic field of the configuration are taken homogeneous is justified by the often slow variation of these quantities in real astrophysical systems, even over scales for which the self-gravity is already relevant (for instance, the primordial cosmological plasma and the ionized intergalactic baryonic component {\cite{Mir}}). Indeed, our study concerns the linear stability of a homogeneous magnetized and self-gravitating plasma, endowed with a finite value of the resistivity coefficient. This latter dissipative feature is here introduced to account for the non-ideal nature of the most commonly observed space plasma. Significant reconnection processes of the magnetic profile are often observed or argued via the interpretation of data from astrophysical configurations. Despite the effects of a finite resistivity coefficients are particularly important in the non-linear regimes, where the establishment of a turbulent profile of the plasma can phenomenologically enforces the resistivity (see for instance the question concerning the so-called \emph{anomalous resistivity} in the configurations of stellar accretion disks  \cite{Bisno01} from the plasma instabilities raised from the streaming of electrons,
and also \cite{1978ApJ...223L..83G,1979ApJ...232..259G,Ghosh:1979ix}), nonetheless, we will show how its presence is crucial already in the linear case, when dealing with the stability properties.

As a first step, we analyze the linear stability for the ideal case, when the resistivity coefficient of the plasma vanishes. In this limit, we essentially reconstruct the Jeans paradigm for the gravitational stability of the plasma structure. The stability out of the plane orthogonal to the constant magnetic field remains still characterized by the same Jeans length obtained originally for the fluid scheme. A relevant new feature emerges, however, in the plane perpendicular to the magnetic force straight lines, where the contribution due to the magnetic pressure affects the equilibrium enhancing the value of the Jeans length by a term corresponding to the square of the Alfv\'en velocity in the plasma. Such additional contribution enters the Jeans length expression on the same footing as the sound speed contribution and therefore its relevance strictly depends on the ratio of the sound speed to the Alfv\'en one. In particular, the greater the Alfv\'en speed is, the larger is the anisotropy in the gravitational collapse, inside and outside the orthogonal plane. We properly describe this effect by following the behavior of a over-dense region during the linear evolution, which is accordingly squeezed on the orthogonal plane. Indeed, in the linear regime, the density contrasts grows without a real gravitational collapse (that takes place essentially in the non-linear stage of the evolution), but the growth is slower on the plane where the magnetic pressure affects the Jeans scale. It is worth noting that, in the perturbation scheme, the magnetic pressure is, despite its name, anisotropic, being provided by the scalar product between the background and the perturbed magnetic field. It is just this feature that introduces the anisotropy in the perturbation evolution. The mode that becomes unstable on the orthogonal plane corresponds to the quasi-stationary one (typical of the slow magnetosonic configuration of a non-gravitating plasma). The presence of gravity alters the nature of this mode, making the system unstable, but with a greater Jeans length with respect to the directions out of this plane, for which the quasi-stationary mode would be absent in the non-gravitational case too.

The gravitational stability analysis is then faced taking into account a non-vanishing resistivity coefficient. In this case, our work outlines how the stability condition is now fixed by the request that the scale length should be smaller than the Jeans value, independently from the direction along which the perturbation mode propagates, as in the standard model. This issue relies on the plasma dispersion relation in presence of a non-zero resistivity. However, the anisotropic evolution of the linear perturbations is not completely removed, since their growth proceeds at a different velocity that depends on the angle between the perturbation wave number and the background magnetic field. This feature is well elucidated in the limit of very small (but non zero) values of the resistivity coefficient, \ie for a very conducting fluid. In fact, although the orthogonal stability is not present when an arbitrarily small value of the resistivity is taken onto account, the time evolution of the perturbations is still affected by a certain dependence on the direction, as far as the resistivity coefficient is not too large. This is due to the small, though real, values that the frequency takes on the orthogonal plane to the magnetic field when the resistivity is very small, in comparison to the other directions.

It is important to remark that in the resistive MHD model proposed in this paper, any other dissipative effects have been neglected (like those driven by viscosity) although in many astrophysical contexts these effects are clearly relevant. Indeed, a correct and more realistic picture of plasma instability mechanism should be properly given by the visco--resistive MHD approach. In fact, the relative magnitude of the viscous and magnetic diffusion rates can be parameterized through the magnetic Prandtl number. For the cosmological plasmas, for instance, the viscous diffusion is more important than resistive diffusion, at recombination and for more recent times. Thus, viscosity dominates on resistivity at all scales of cosmological interest and becomes relevant (\emph{i.e.}, the associated frequency becomes equal to the Alfv\`en frequency) at recombination for scales of the order of $10$ Kpc comoving or less. Contrarily, the corresponding scale for the resistivity is very small and it is under the scales relevant for cosmology. However both these ``critical'' scales depend on time and, therefore, the viscosity always dominates over resistivity, but both dissipative effects can indeed be neglected when addressing the propagation of Alfvén waves\cite{Lattanzi:2011hu}.

Summarizing, our study outlines the role that the plasma nature of the system plays in the gravitational instability dynamics. We generalize the concept of Jeans length and outline the anisotropic dynamics induced by the magnetic pressure, both in the ideal and in the resistive case. The obtained results acquire particular interest when they are treated as initial conditions for the non-linear and turbulent collapsing dynamics. The paper is organized as follows. In Sec.\il\ref{Sec:uno}, we provide the fundamental equations for the resistive MHD equilibrium and develop the perturbation scheme. The resulting differential system is combined in order to obtain a single ordinary differential equation for the fluid velocity, which gives the dispersion relation describing the system stability. Then, we discuss the zero-resistivity case in Sec.\il\ref{Sec:zere} and the case with finite resistivity Sec.\il\ref{Sec:res}. In both cases we analyze the evolution of the density contrast in momentum and coordinate space, in order to outline the growth of the anisotropy. Finally, discussion and conclusions follow in Sec.\il\ref{Sec:fi}.

\section{Resistive perturbative MHD dynamics}\label{Sec:uno}
\subsection{Basic equations}
In this Section we recall the basic equations of resistive MHD. The mass conservation and the configuration of the Newtonian gravitational field are described by the continuity and Poisson equations
\begin{equation}
\pt\rho+\vna\cdot\rho\vec{v}=0\;,
\qquad\qquad
\lapl\Phi-4\pi G\rho=0\;,
\label{eq:contpos}
\end{equation}
respectively, where $\rho$ is the mass density, $\vec{v}$ is the velocity field, $\Phi$ is the gravitational potential and $G$ is Newton constant. Moreover, the single-fluid dynamics of the plasma is described by the Euler equation in the presence of a magnetic field $\vec{B}$, \emph{i.e.},
\begin{equation}\label{eq:Euler}
\rho\pt\vec{v}+\rho(\vec{v}\cdot\vna)\vec{v}+\vna P+\rho\vna\Phi
-(\vna\times\vec{B})\times\vec{B}/4\pi=0\;.
\end{equation}
The electromagnetic interaction is summarized by the Maxwell equations for the electric ($\vec{E}$) and magnetic fields
\bea
\pt\vec{B}+c\,(\vna\times\vec{E})=0\;,&\label{eq:Faraday}\\
4\pi\,\vec{J}-c\,(\vna\times\vec{B})=0\;,&\label{eq:Ampere}\\
\vna\cdot\vec{B}=0\;,&\label{eq:Gauss}
\eea
and by the Ohm law for the density current $\vec{J}$
\begin{equation}
\vec E +\frac{\vec v \times \vec B}{c}=\eta \vec J\;. \label{eq:elforcebal}
\end{equation}
Combining Eqs. \reff{eq:Faraday}, \reff{eq:Ampere}  and \reff{eq:elforcebal} we get the fundamental relation for the evolution of the magnetic field:
\begin{equation}
\pt\vec{B} = \vna\times(\vec{v}\times\vec{B})+\etab\lapl\vec{B}\;,
\label{eq:base2}
\end{equation}
where we have introduced the diffusion coefficient $\etab\equiv\eta c^2/4\pi$. For our analysis, the dynamics is summarized by \erefs{eq:contpos}, \reff{eq:Euler} and \reff{eq:base2}.

\subsection{Perturbation scheme}\label{Sec:sue}
In order to analyze the implications that the physics of a resistive plasma has on the structure formation mechanism, we now follow the standard perturbation approach, as in the Jeans model \cite{Libro}. In this respect, we consider a small perturbation around a static and uniform solution (indicated by the subscript $0$). Thus, we assume $\rho=\ro+\ru$, with $\ru\ll\ro$ (the same is also valid for $P$, $\vec v$, $\Phi$ and $\vec B$) where the background is characterized by constant $\ro$, $\po$, $\phio$, $\bo$ and $\vo=0$ \cite{Mir,Wei}. We follow the standard procedure that takes this as a solution of the above equations even if the Poisson equation is not actually satisfied (unless $\ro=0$). This fact, that corresponds to neglect the effects of gravitation in the unperturbed solution, is known as the ``Jeans swindle'' (see for example \cite{Mir}).

In this picture, the system describing the first-order perturbative dynamics reads now
\bea
\pt\ru+\ro\vna\cdot\vu=0\;,&\label{eq:contpospert}\\
\lapl\phiu-4\pi G\ru=0\;,&\label{poisson-pert-static}\\
\ro\p_t\vu+v_s^2\vna\ru+\ro\vna\phiu
-(\vna\times\bu)\times\bo/(4\pi)=0\;,&\label{eq:Eulerpert}\\
\pt\bu-\etab\lapl\bu+\bo(\vna\cdot\vu)-(\bo\cdot\vna)\vu=0\;,&\label{eq:froz}\\
\vna\cdot\bu=0\;,&
\eea
where the pressure and density perturbations are related through the adiabatic sound speed, \emph{i.e.}, $\pu=v^2_s\ru$. Introducing  the dimensionless magnetic fluctuation $\b\equiv\bu/B_0$ (where $\bo=\hbo\,B_0$), we  consider perturbations $\psi$ of the following form
\begin{align}
\psi(\rv,t)=\tilde{\psi}_1(t)e^{-i \kv\cdot\rv}\;,
\end{align}
where $\psi_1$ stands for $[\ru,\;\vu,\;\b,\;\Phi_1]$ and $\vec q=q\,\hq$ denotes the wave-number (in the following we will drop the tilde for the sake of simplicity). This way, the corresponding evolution equations in the harmonic space can be simply obtained by the substitution $\vna \to i\kv$. A further simplification can be obtained by decomposing $\vu$ as $\vu=\vup\,\hq+\vuo$ (where $\vuo\cdot\hq=0$), and by introducing the following variables
\begin{align}\label{scalar-var}
\delta=\ru/\ro\;,\qquad\qquad
\theta=i(\kv\cdot\vu)\;,\qquad\qquad
\bb=\b\cdot\hbo\;,\qquad\qquad
\vb=iq(\vuo\cdot\hbo)\;.
\end{align}

In this scheme, the perturbative dynamics results to be described by the following system which involves only scalar quantities
\begin{subequations}\label{master-system}
\begin{align}
\dot{\delta}+\theta=0&\;,\\
\dot{\theta}-\omega_0^2\delta-\oma\bb=0&\;,\\
\dot{\bb}+\theta(1-\mu^2)-\mu\vb+\ome\bb=0&\;,\\
\dot{\vb}+\mu\oma\bb=0&\;,
\end{align}
\end{subequations}
where we have defined
\begin{align}
\mu\equiv\hbo\cdot\hq\in [0,1]\;,\qquad
\omega_0^{2}\equiv v_s^{2}q^2-4\pi G\ro\;,\qquad
\omega_A \equiv \displaystyle\frac{B_0^2}{ 4\pi\rho_0} q\equiv v_A q\;,\qquad
\omega_\eta\equiv\bar\eta q^2\;.\qquad
\end{align}
The system \reff{master-system} can be now reduced to a unique equation for the variable $\theta$
\begin{align}\label{master-theta}
\theta^{(4)}+\etab q^{2}\;\theta^{(3)}+
(\omo+\oma)\;\theta^{(2)}+\etab q^{2}\omo\;\theta^{(1)}+
\mu^{2}\oma\omo\;\theta=0\,,
\end{align}
where $\theta^{(\ell)}$ denotes the $\ell^{th}$ derivative with respect to time. Since \reff{master-theta} is a linear, fourth-order, ordinary differential equation with constant coefficients, the explicit solution is of the form $\theta(t)\propto e^{i\omega t}$. The same consideration stands for all the variables of \eref{scalar-var} also. This way, we get the following algebraic equation for the variable $\omega$
\be\label{eq:disp}
\omega^4+i a_3 \omega^3+a_2 \omega^2+i a_1 \omega+a_0=0\;,
\ee
where we have set
\begin{align}\label{abdefi}
a_0\equiv\mu^{2}\oma\omo\;,\qquad\quad
a_1\equiv\ome\omo\;,\qquad\quad
a_2\equiv-\,(\oma+\omo)\;,\qquad\quad
a_3\equiv-\ome\;.
\end{align}

Since all the quantities admit the same time dependence, \textit{e.g.}, $\delta\propto e^{i\omega t}$, the gravitational stability is determined by the sign of $\omega^2$, \emph{i.e.}, if the imaginary part of $\omega$ results to be negative. A solution with $\omega^2>0$ ($<0$) corresponds to an oscillating (exponential) perturbation, while for $\omega=0$ one gets a static density contrast. In the following, we will focus on the analysis of the solutions of \eref{eq:disp} discussing in some details the different cases.

\section{Zero-resistivity case: analytical and numerical analysis}\label{Sec:zere}
Let us start by considering the ideal MHD limit. When $\etab=0$, \eref{eq:disp} admits the following solutions:
\bea\label{takreal}
\omega^2_\pm(q)=\tfrac{1}{2}\;
\Big[\;(\omo+\oma)\pm\sqrt{(\omo+\oma)^2-4\mu^2\oma\omo}\;\Big]\;.
\eea
We recall that $\omega_0^2$ can be positive or negative, depending on the relative strength of the pressure and gravitational force. In fact, the standard Jeans picture is recovered when $\oma=0$: in this case, the solution is simply $\omega^2=\omo$ and the oscillating ($\omega^2>0$) and collapsing ($\omega^2<0$, {taking the negative solution}) modes are separated by the critical (Jeans) wave-number
\begin{equation}
q_J \equiv \sqrt{\frac{4\pi G \ro}{v^2_s}}\;.
\end{equation}

In the presence of a magnetic field ($\oma\ne0$), it can be shown that
\begin{equation}
\rm{sign}(\omega^2_\pm)=
\left\{
\begin{array}{ll}
\rm{sign}(\omo+\oma)\quad &\mu = 0
\;, \\[0.2cm]
\rm{sign}(\omo) \quad & \mu\ne0
\;.
\end{array}
\right.
\end{equation}
Here, the relation for $\mu=0$ is evaluated by considering the $\omega_+$ solution since, in this case, $\omega_-^{2}$ results always positive yielding an oscillating density contrast evolution (and vice versa for $\mu\neq0$). This implies that, when the perturbation wave-vector lies in the plane orthogonal to the background magnetic field ($\mu =0 $), the magnetic pressure can induce stable perturbations that would otherwise collapse in the absence of a $B$-field. In fact, even if $\omo<0$ (indicating an instability according to the standard Jeans criterion), nevertheless $\omo+\oma >0$ if $\omega_{\ti{A}}$ is sufficiently large. The corresponding critical wave-number $q_c$ is readily found to be
\begin{equation}
q_{c} = \sqrt{\frac{4\pi G \ro}{v^2_s+v_{\ti{A}}^2}} \qquad (\mu=0)\;.
\end{equation}
In particular, perturbation modes with $q>q_c$ are stable, while those with $q<q_c$ grow exponentially. Since $q_c$ is always smaller than $q_J$, the presence of the magnetic field increases the stability of the system. By other words, the critical wave-number is simply obtained by replacing, in the expression for the Jeans wave-number, the sound speed with the effective speed $v_{eff}^2=v_s^2+v_{\ti{A}}^2$. On the other hand, as soon as the perturbation is not exactly perpendicular to the background field ($\mu \ne 0$), the stability of the perturbation is dictated by the standard Jeans criterion, \emph{i.e.}, $q_c = q_J$.

Concluding this section, let us now introduce for completeness the standard Jeans length and the magnetic critical length defined as
\be
\lambda_J=\sqrt{\frac{\pi v_s^2}{G \rho_0}}\;,\qquad\qquad
\lambda_{JM}=\sqrt{\frac{\pi(v_s^2+v_{\ti{A}}^2)}{G \rho_0}}\;,
\ee
respectively. We recall that the second quantity defines the stability in the plane orthogonal to the background magnetic field, \emph{i.e.}, $\mu=0$.

\subsection{Time evolution of the density contrast (Fourier space)}
\begin{figure}[!ht]
\centering
\includegraphics[width=0.45\hsize]{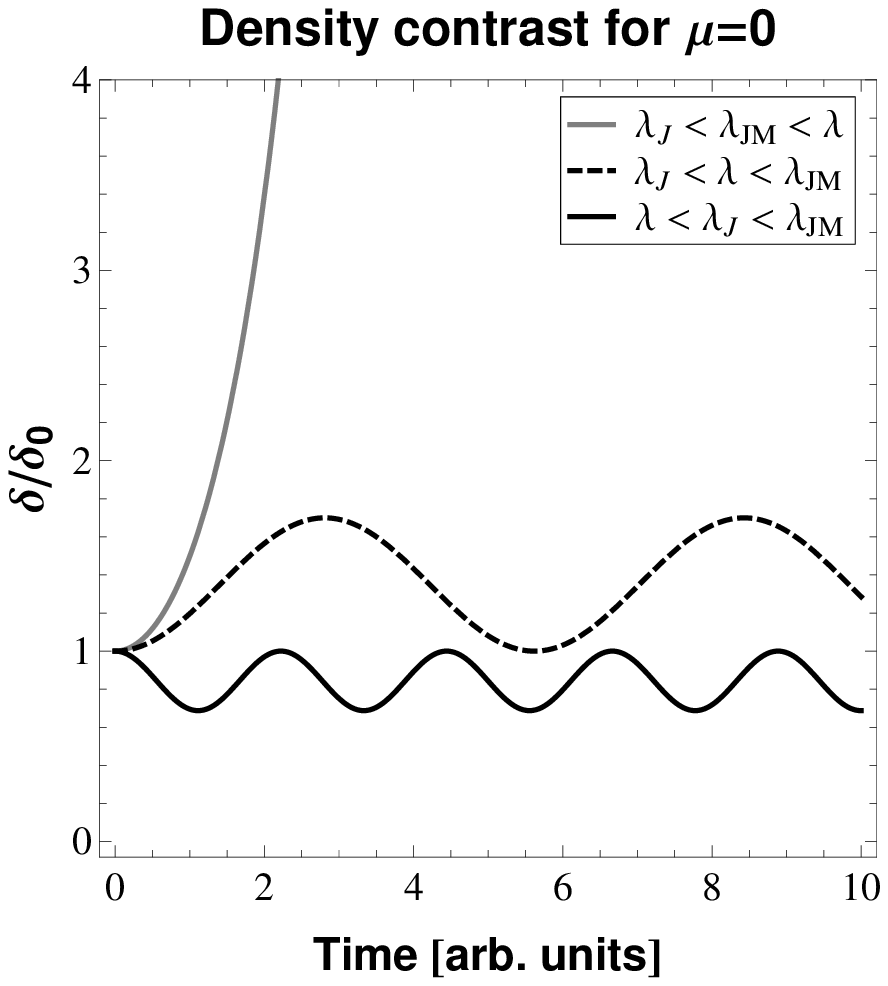}
\includegraphics[width=0.45\hsize]{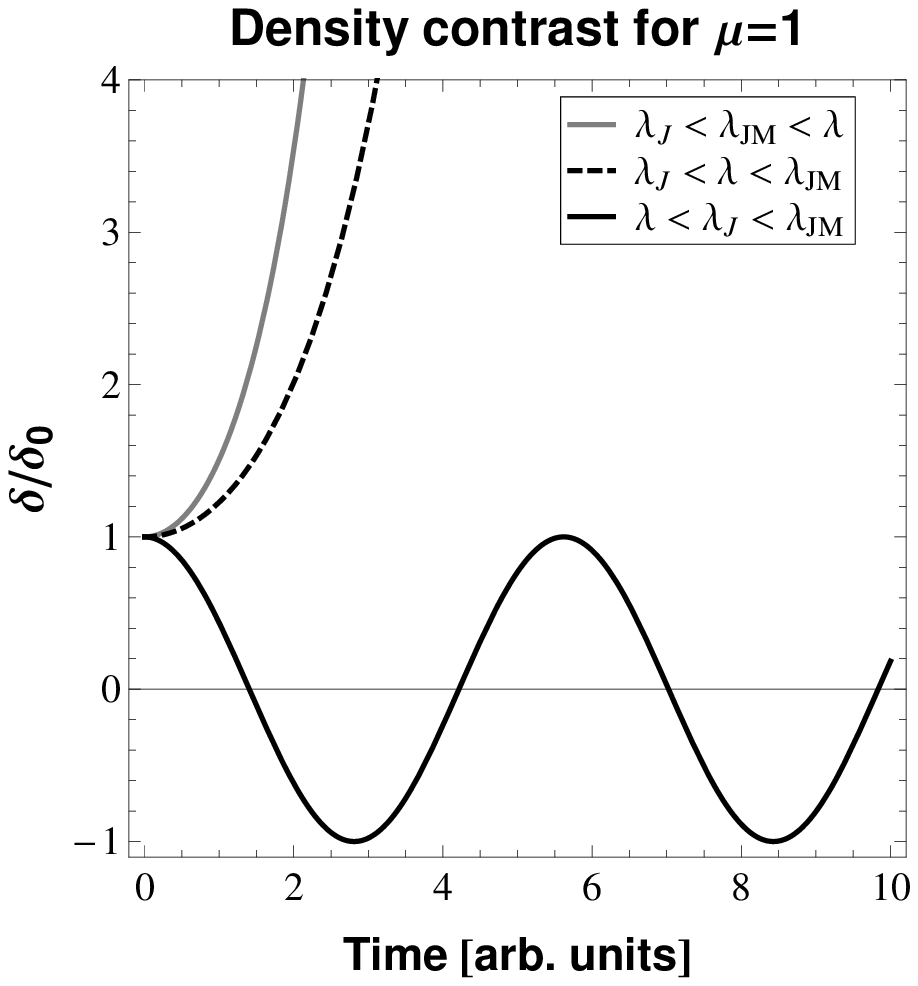}
\caption{Plot of the ratio $\delta(t)/\delta_0$ as function of the time $t$, for $\mu^2=0$ (left panel) and $\mu^2=1$ (right panel).}
\label{prima}
\end{figure}

In what follows, {we numerically} integrate the system \reff{master-system}, for $\ome=0$, in order to characterize the behavior of the density contrast. The initial conditions are set as $\delta(t_0)=\delta_0$, $\bar{v}(t_0)=0$, $\bb(t_0)=0$ and $\dot{\delta}|_{t_0}=0$. In Figure~\ref{prima}, we show the time evolution for perturbations with different wave-number, in the directions orthogonal and parallel to the magnetic field, respectively.
It is evident that fluctuations with $\lambda_J<\lambda<\lambda_{JM}$, which would result unstable according to the standard Jeans criterion, are stabilized in the direction orthogonal to the background magnetic field.

\subsection{Time evolution of the density contrast (coordinate space)}\label{sunseinfani}
From the previous analysis, one would expect the gravitational collapse to be anisotropic. In this sense, in order to outline the nature of the anisotropy induced by the magnetic field in the physical space, we now consider an initial Gaussian over-dense region in real space, \emph{i.e.},
\begin{align}
\delta(\vec{r},t=0)=\delta_c\; e^{-r^2/2\sigma^2}\;,
\label{ingauss}
\end{align}
where the background magnetic field $\vec{B}_0$ is directed along the $y$-axis. The corresponding initial over-density in Fourier space is still Gaussian {(}in this Section, we restore the tilde in order to denote quantities in harmonic space{)}:
\be
\tilde{\delta}(\vec{q},t=0)=\frac{1}{(2\pi)^{3/2}}
\int\delta(\vec{r},t=0)e^{i\vec{q}\cdot\vec{r}}d^3\vec{r}=
\tilde{\delta}_c\; e^{-q^2/2\tilde{\sigma}^2}\;,
\ee
and, evolving such over-density in time, \ie $\tilde{\delta}(\vec{q},t)$, we are able to (numerically) transform back to real space getting the profile of the density contrast at a given time.
\begin{figure}[!ht]
\centering
\includegraphics[width=0.5\hsize]{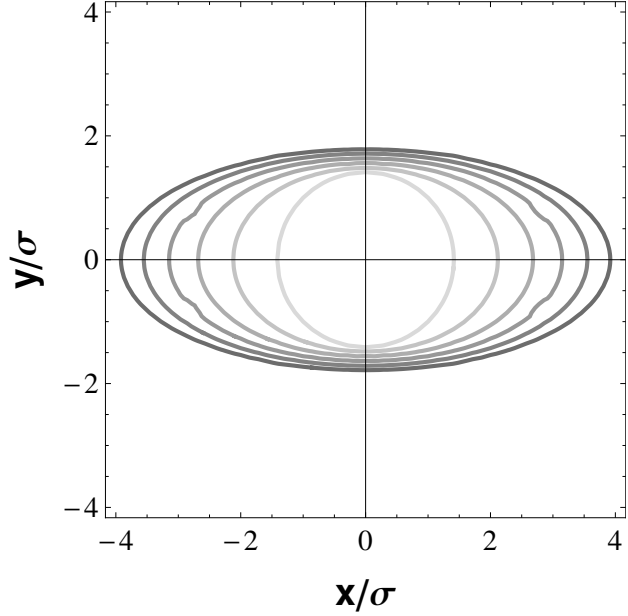}
\caption{Plot of the isolines where $\delta=\delta_c/e$, in the plane $[x/\sigma,\;y/\sigma]$ ($z=const$) at different times (light gray=earlier, dark gray=later).
}
\label{profbi}
\end{figure}
Taking $v^2_A = 10v^2_s$ ($\lambda_{JM}\sim3\lambda_J$) and $\sigma=2\lambda_J$ (so that $\lambda_J<\sigma<\lambda_{JM}$), in  Figure \ref{profbi}, we show the contour line of the density contrast where $\delta=\delta_c/e \simeq 0.37 \delta_c$ at different times. As a result, we find that the density contrast isolines tend to ``squeeze'' (for increasing time) along the direction orthogonal to the background magnetic field.

This issue shows how the presence of the constant magnetic field confining the plasma can play an important role for the perturbation evolution since the collapse in the real space favors the formation of squeezed structures. Indeed, our analysis is limited to the linear evolution and we cannot make prediction about the final fate of such a collapsing profile, but it is remarkable that the initial conditions for the non-linear phase of this process results to contain an intrinsic anisotropic feature.

\section{Resistive case}\label{Sec:res}

Let us now discuss in some details the stability properties of the the pure resistive MHD case. By setting $\eta\neq0$ in \eref{eq:disp}, the four solutions for $\omega$ write
\begin{align}\label{master-sol}
\omega^{(\pm)}_{\mp}=-\frac{i a_3}{4}\pm
\frac{1}{2}\sqrt{\phi-\frac{2 a_2}{3}-\frac{a_3^2}{4}}\mp
\frac{1}{2}\sqrt{-\phi-\frac{4 a_2}{3}-\frac{a_3^2}{2}-
\frac{i\left(-8a_1+4a_2a_3+a_3^3\right)}{2\sqrt{4\phi-\frac{8a_2}{3}-a_3^2}}}\;,
\end{align}
where we have defined the following constants:
\begin{align}
\phi\equiv&\frac{2^{1/3} \gamma}{3 p^{1/3}}+\frac{p^{1/3}}{3 (2)^{1/3}}\;,
\qquad\qquad p\equiv\sigma +\sqrt{\sigma^2-4 \gamma^3}\;,\\
\gamma\equiv&\left(12a_0+a_2^2+3a_1a_3\right)\;,
\qquad\sigma\equiv 2 a_2^3-27 a_1^2+9 a_2 a_1 a_3-9 a_0 \left(8 a_2+3 a_3^2\right)\;.
\end{align}
In what follows, we consider oscillating and growing (or decreasing) perturbations separately and, for the sake of simplicity, we split $\omega$ into its real and imaginary parts $x$ and $y$,
\begin{align}
\omega=x+iy\;,
\end{align}
where $x$ and $y$ are both real.

At a first instance, it is possible to show that real (non-zero) solutions \reff{master-sol} exist only in the case $q>q_J$ ($\omo\geqslant0$) and for $\mu=1$. Recalling that $\delta\sim e^{i\omega t}$, real frequencies $\omega^{2}>0$ correspond to the oscillating modes and, in this case, the solution results to be $\omega=\pm\omega_0$ and does not depend on the coefficient $\ome$. From this result, one can argue that pure oscillating perturbations are admitted only in the plane parallel to the background magnetic field, \ie $\mu=1$, and for $q>q_J$. In this regime, the modes are not affected by the presence of resistivity.

Let us now turn our attention on the case of growing or decreasing perturbations. For scales smaller than the Jeans length and for $\mu\neq1$, the imaginary part $y$ of $\omega$ is always different from zero and, as we will show in the next Section, only solutions with $y>0$ are {admitted} (the regime $y<0$ {does not occur}). This fact results in a damping of the density contrast for $q>q_J$. On the other hand, in the case of larger scales, \ie $q<q_J$, there are no oscillating perturbations and we find at least one growing exponential solution with $y<0$ and $x=0$. In this sense, we can conclude that, also in the presence of resistivity, the basic Jeans instability criterion is confirmed and the critical wave-number which discriminates the gravitational collapse results, in turn, to be $q=q_J$. For a better comprehension of these results, we now discuss in some details the different specific cases. Since the condition to have a growth of the density contrast is $y < 0$, in the following we do not consider purely real solutions.

\subsection{Below the Jeans length: $q>q_J$}\label{B}

\paragraph{Imaginary frequencies: $\omega=iy$}
We consider at first the purely imaginary solution where $\omega=i y$. We are interested in particular in defining the sign and the expression of $y$. It is convenient to introduce here the following notation
\be
y^{\pm}_{(\mp)}\equiv\frac{\ome}{4}\pm\frac{1}{4} \sqrt{\ome^2-\frac{8}{3} \left(\omo+S^2 \ome^2\right)+4 \Phi }\mp\frac{\sqrt{\ome^2-\frac{8}{3} \left(\omo+S^2 \ome^2\right)-2 \Phi -\frac{\ome \left(4 \omo+\left(1-4 S^2\right) \ome^2\right)}{\sqrt{\ome^2-\frac{8}{3} \left(\omo+S^2 \ome^2\right)+4 \Phi }}}}{2 \sqrt{2}}\;,
\ee
and
\be
\bar{y}^{\pm}_{(\mp)}\equiv\frac{\ome}{4}\pm\frac{1}{4} \sqrt{\ome^2-\frac{8}{3} \left(\omo+S^2 \ome^2\right)+4 \Phi }\mp\frac{\sqrt{\ome^2-\frac{8}{3} \left(\omo+S^2 \ome^2\right)-2 \Phi +\frac{\ome \left(4 \omo+\left(1-4 S^2\right) \ome^2\right)}{\sqrt{\ome^2-\frac{8}{3} \left(\omo+S^2 \ome^2\right)+4 \Phi }}}}{2 \sqrt{2}}\;,
\ee
where the Lundquist number $S\equiv {\omega_{\ti{A}}}/{\ome}={v_{\ti{A}} }/{\etab q}
={4\pi v_{\ti{A}}}/{\eta qc^2}$  characterizes the properties of the plasma. In fact, at fixed scale, high values of $S$ indicate strongly conducting plasmas. Let us also set
\be
\Phi \equiv\frac{P^{1/3}}{3 (2)^{1/3}}+\frac{2^{1/3} \Gamma }{3 P^{1/3}}\;,\qquad\qquad
P\equiv\Sigma +\sqrt{(\Sigma )^2-4 (\Gamma )^3}\;,
\ee
\be
\Sigma \equiv 27 \omo \ome^2\left(\omo +\mu^2 S^2 \ome^2\right)-\left(\omo+S^2 \ome^2\right)\left[9 \omo \ome^2 \left(1+8\mu^2 S^2 \right) -2 \left(\omo+S^2 \ome^2\right)^2\right]\;,
\ee
\be
\Gamma \equiv 3 \omo \ome^2\left( 4 \mu^2 S^2 -1\right)+\left(\omo+S^2 \ome^2\right)^2\;.
\ee
Finally we name
\be
y_1\equiv y_-^-,\quad y_2\equiv y^-_+,\quad y_3\equiv \bar{y}^+_-,\quad y_4\equiv \bar{y}_+^+.
\ee
In the regime $q>q_J$, purely imaginary solutions (when these exist) of Eq.\reff{eq:disp}  belong to the set $\{i y_1,\,i y_2,\, i y_3,\, iy_4\}$. It can be verified that all the $y$'s are always greater than zero. This implies that, for $q>q_J$, purely imaginary solutions always describe perturbations that are exponentially damped away.

\paragraph{Complex frequencies: $\omega=x+iy$}
In this section we focus on the most general case $x\neq0$ and $y\neq0$.
It turns out again that $y$ is always positive, while $x=\pm x_s$ where
\be
x_s\equiv\sqrt{\frac{\omo \ome-2 \left(\omo+S^2 \ome^2\right) y+3 \ome y^2-4 y^3}{\ome-4 y}}\;.
\ee
A very interesting case is given by the value $y=\ome/4$, where $x^2=x^2_{\pm}$, \textit{i.e.},
\be
x^2_{\pm}\equiv\frac{1}{16} \left[8 \omo+\left(8 S^2-3\right) \ome^2\pm2 \sqrt{16 \omega_{\ti{0}}^4+32 \omo (1-2 \mu^2) S^2 \ome^2+\left(16 S^4-16 S^2+3\right) \ome^4}\right]\;,
\ee
\be\label{otilde}
\omo=\tilde{\omega}_0^{2}\equiv\ome^2 \left(4 S^2-1\right)/4\;.
\ee
As soon as we are considering the case $\omo>0$,  from Eq.\il(\ref{otilde}) we must restrict our analysis to the case $S>1/2$. {We can identify the following different regions for the parameters:
\begin{itemize}
\item[-]
$3\sqrt{2}/8<S\leq\sqrt{5}/4$ and $0\leq\mu^2<\mu^2_o$: it results $x^2=x^2_+$;
\item[-]
$S>\sqrt{5}/4$ and $0\leq\mu^2\leq\mu^2_o$: it results $x^2=x^2_+$;
\item[-]
$S>\sqrt{5}/4$ and $\mu^2_o<\mu^2<\tilde{\mu}^2_o$: it results $x^2=x^2_{\pm}$;
\item[-]
$\mu^2=\tilde{\mu}^2_o$: it results $x^2=x^2_{+}$;
\end{itemize}
where we have defined
\bea
\mu_o^2&\equiv& \frac{48 \omo+\left(3-16 S^2\right) \ome^2}{256 \omo S^2}\;,
\qquad\qquad
\tilde{\mu}_o^2\equiv\frac{\ome^4(3-16 S^2)+16 \left(\omo+S^2 \ome^2\right)^2}{64 \omo \ome^2 S^2}\;.
\eea}

We can conclude that, in this parameter domain, the only significant information concerns the behavior of the frequency characterizing the oscillations of the perturbed quantities and, therefore, we can rule it out from the study of the gravitational stability.

\subsection{Above the Jeans length: $q<q_J$}\label{C}
In this Section, we consider the case $q<q_J$. We find that there are no purely oscillating perturbations ($y=0$), \emph{i.e.}, that the time dependence of the perturbation always contains an exponential part. In particular, there is always one and only one exponentially growing mode with $y=y_1<0$ and $x=0$ and at least one exponentially damped mode with $y>0$ and $x=0$. {This scenario} is explored in detail in the following.
\paragraph{Imaginary frequencies: $\omega=iy$}
For $q<q_J$, there is always one (and only one) solution of Eq.\reff{eq:disp} with imaginary part $y=y_1<0$. There is also at least another purely imaginary solution of Eq.\reff{eq:disp}, having $y>0$ and  belonging to the set $\{ i y_2,\, i y_3,\, iy_4\}$.
The existence of other purely imaginary solutions will depend in general on the value of the parameters $\mu$, $\omo$ and $S$. However, this result shows that, above the Jeans length, an unstable mode is always present.

\paragraph{Complex frequencies: $\omega=x+iy$}
We focus now on the case $x\neq0$ and $y\neq0$. The frequency $\omega=x+iy$ admits the solutions $x=\pm x_s$ (for $y\neq \ome/4>0$, decreasing) and $y=\tilde{y}_2$ where
\bea
\tilde{y}_2&\equiv&\frac{\ome}{4}+\frac{\sqrt{6 (2)^{2/3} \nu +12 (2)^{1/3} \nu ^{1/3} \upsilon -3 \left[8 (S^2 \ome^2+\omo)-3 \ome^2\right] \nu ^{2/3}}}{12 \nu ^{1/3}}\;,
\eea
with
\bea
\nu &\equiv&\delta +\sqrt{\delta ^2-4 \upsilon ^3}\;,\\
\delta &\equiv&2 \omega_{\ti{0}}^6 + 6 \omega_{\ti{0}}^4 (3 + (1 - 12 \mu^2) S^2) \ome^2 +
 3 \omo S^2 (-3 + 2 S^2 + \mu^2 (9 - 24 S^2)) \ome^4 + 2 S^6 \ome^6\;,\\
\upsilon&\equiv&S^4 \ome^4+2 S^2 \ome^2 \omo(6  \mu^2+1)+\omo \left(\omo-3 \ome^2\right)\;.
\eea
The solution $\tilde y_2$ is founded to be always positive. Thus we can conclude that for $q<q_J$, complex solutions of Eq. \reff{eq:disp}, when they exist, always correspond to oscillating decreasing modes. Together with the result for the purely imaginary case, this implies that above the Jeans length, there is always one and only one
unstable mode, independently on the values of the others parameters.

\section{Limiting cases of small resistivity and anisotropy}
The comparison between the resistive and non-resistive case underlines a difference that may seem odd at first; namely, the presence of an even infinitesimally small resistivity seems to completely erase the stabilizing effect (in the orthogonal plane) of the magnetic field, since in the resistive case we recover the standard, non-magnetic Jeans criterion for gravitational instability. One could then conclude that the support effect of the magnetic field is never actually realized in practice, since an infinitely conducting fluid is obviously just a mathematical idealization. However, an analysis of the limiting case of ``very strong" magnetic field and ``small'' resistivity shows that this is not the case, since the timescale for collapse on the orthogonal plane is much larger than outside that same plane, where it recovers the standard value, only dictated by the action of gravity. The orthogonal timescale becomes actually infinite when $\eta$ is exactly 0, reproducing the result of the non-resistive case described in the previous section.

We then consider the four solutions of Eq.\il(\ref{master-sol}) in the limit of strong magnetic field and small resistivity, \textit{i.e.}, $\oma\gg\ome^2$ and $\oma\gg|\omo|$. By other words, we take the Alfv\'en timescale to be much shorter than the resistive diffusion and standard gravitational collapse timescales.
Thus we write $\omo = - \epsilon^2 \oma$ with $\epsilon \ll 1$ (we choose the minus sign because we are interested in the Jeans-unstable solution), and we perform a series expansion of the frequency with respect to the small parameters $\epsilon$ and $\epsilon' = \ome/\omega_a=1/S$.
We focus on the only Jeans-unstable mode, whose frequency in this limit reads (we only keep leading terms):
\be
\omega = \left\{
\begin{array}{ll}
\displaystyle{-i\, \epsilon^2\epsilon'\omega_A \simeq i \frac{\omo \ome}{\oma}}, & \qquad (\mu = 0)\,; \\[0.3cm]
\displaystyle{- i \epsilon \omega_A \simeq - i \sqrt{|\omo|}}, & \qquad(\mu = 1),
\end{array}
\right.
\ee
in the direction orthogonal and parallel to the magnetic field, respectively. The mode is clearly unstable ($\mathrm{Im}(\omega)<0$) in both directions. However, the timescale over which the over-density grows is very different in the two cases. In the parallel direction, the leading term is of first order in the expansion parameters (and we recover the standard Jeans result $\omega^2=\omo$); in the orthogonal direction, it is instead of \emph{third} order in the expansion parameters. In particular, the frequency in the latter case is suppressed by a factor $\epsilon/S$ with respect to the former, and actually becomes zero (\ie the mode is stable) for an infinitely conducting plasma ($S=\infty$). If the timescale over which orthogonal perturbations grow is much larger than the time interval over which the system is observed, these perturbations can be considered stable for all practical purposes. In any case, they grow much slower than their parallely-oriented counterparts, and this open the possibility to observe an anisotropic collapse like in the case of non-resistive plasmas.

\subsection{Time evolution of the density contrast}
Let us now analyze the time evolution of the density contrast exploring the anisotropy of the gravitational collapse for different values of the resistivity $\eta$. In this sense, we consider an initial Gaussian over-density in real space, \ie of the form \reff{ingauss}, and a background magnetic field $B_0$ directed along the $y$-axis (as assumed for the zero resistivity case). The corresponding initial over-density in Fourier space has still a Gaussian profile. We evolve the Fourier over-density in time, as shown in Section.\il(\ref{sunseinfani}), in order to get $\tilde{\delta}(\textbf{q};t)$ and then (numerically) transform back to real space. Moreover, we set $v^2_A = 10v^2_s$ ($\lambda_{JM}\sim3\lambda_J$) and $\sigma=2\lambda_J$ (so that $\lambda_J<\sigma<\lambda_{JM}$).

\begin{figure}[!ht]
\centering
\includegraphics[scale=.5]{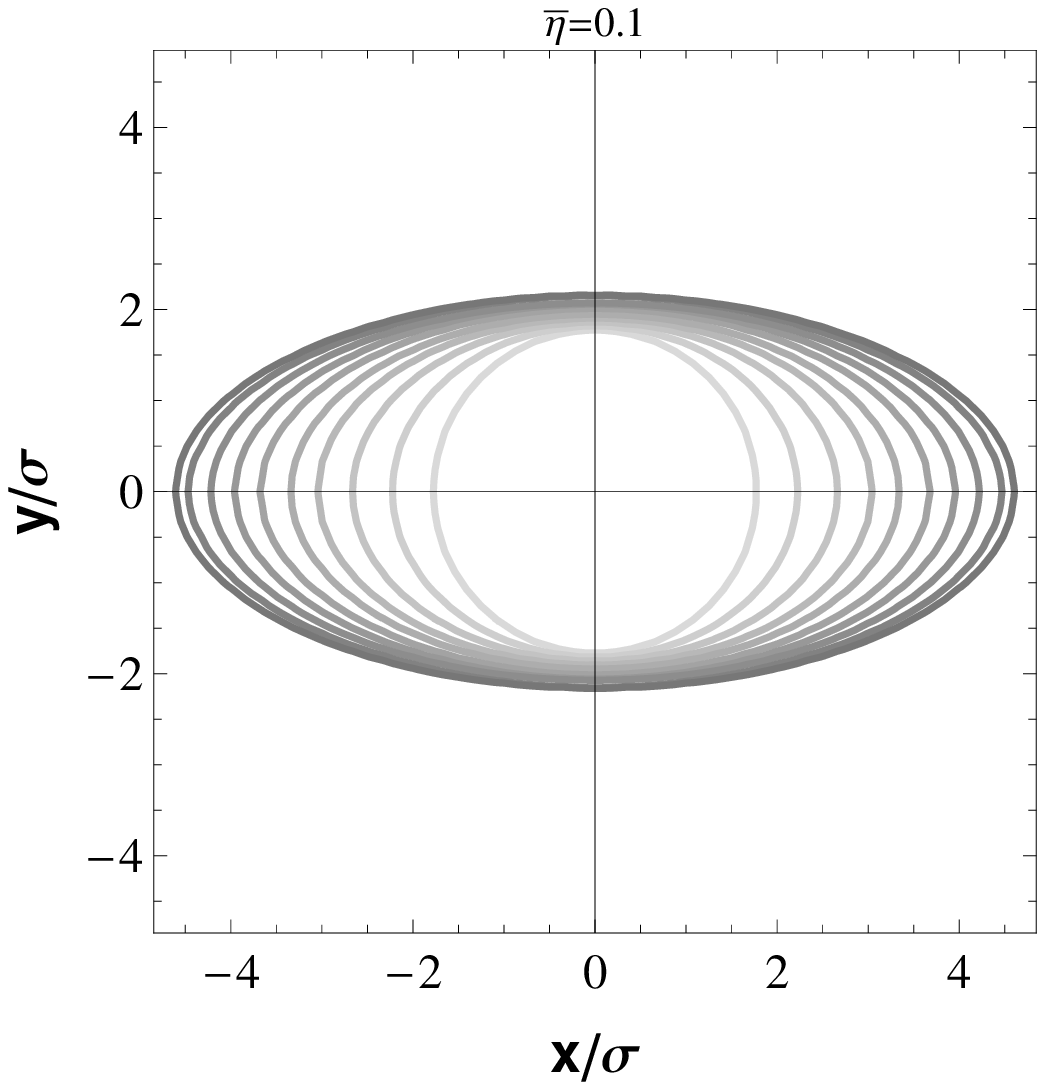}
\includegraphics[scale=.5]{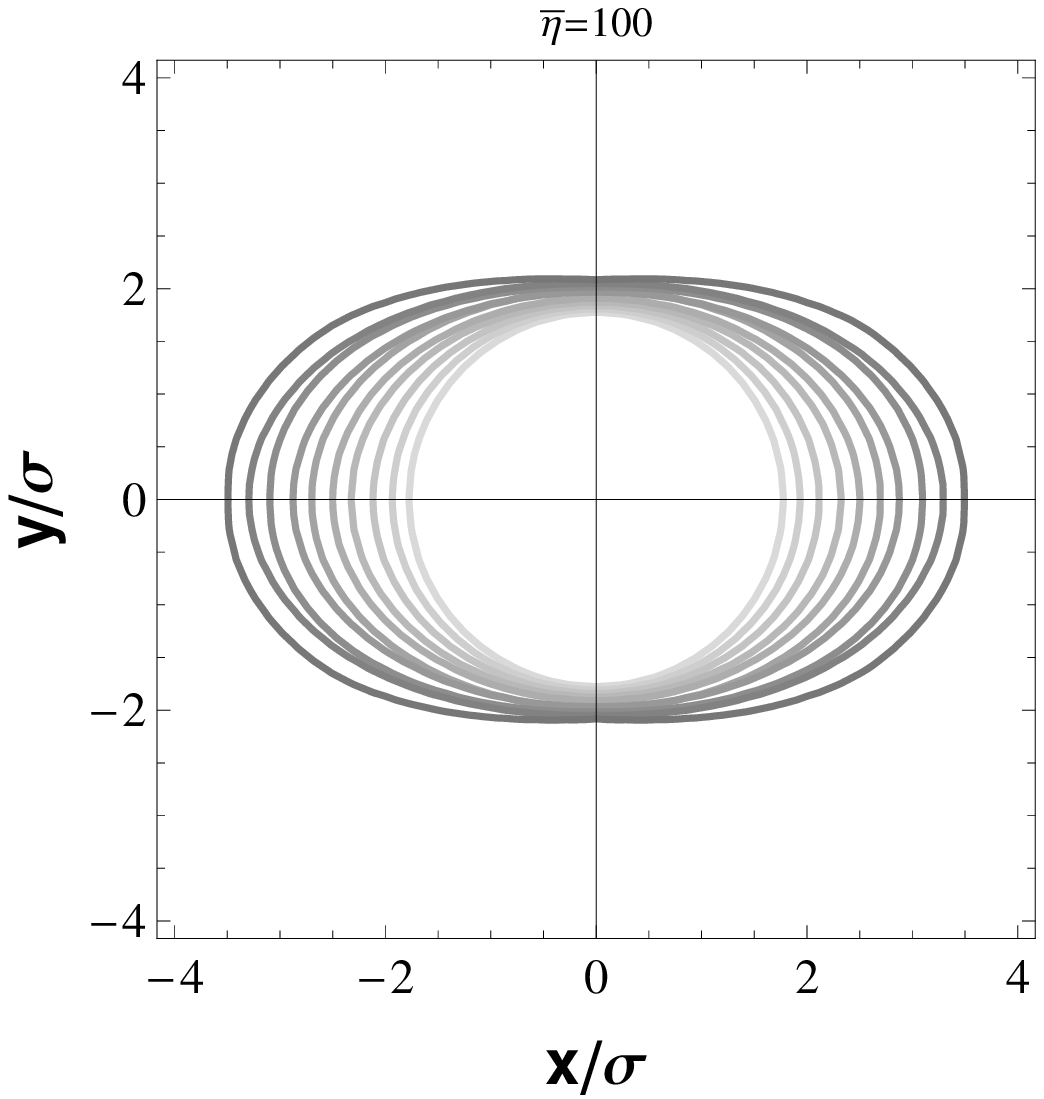}
\includegraphics[scale=.5]{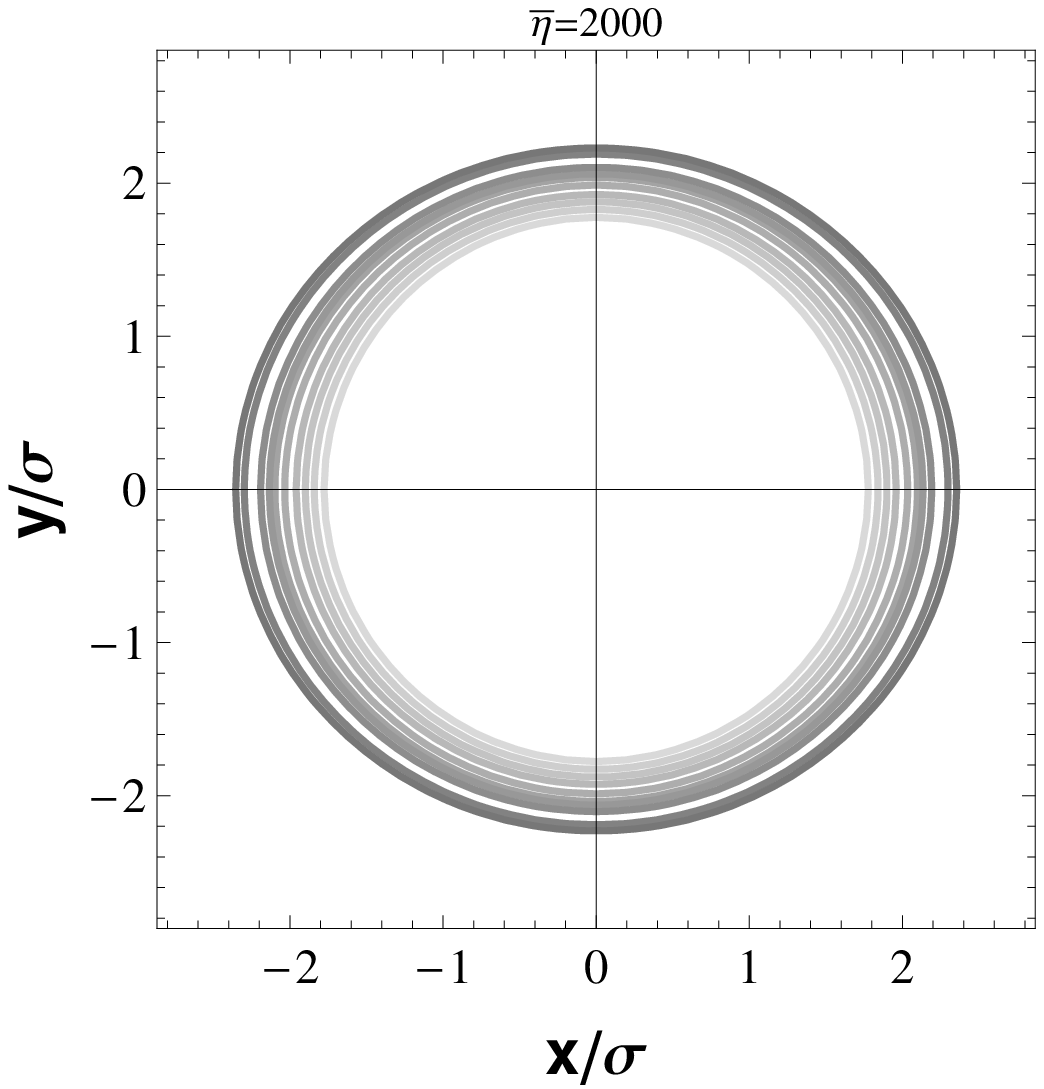}
\caption{Equal density contours corresponding to $\delta=0.99 \delta_c$ at different
times (light gray=earlier, dark gray=later). We take $v^2_A = 10v^2_s$ ($\lambda_{JM}\sim3\lambda_J$) and $\sigma=2\lambda_J$. From left to right, the Lundquist number at the Jeans scale is $S(q=q_J) = \{30,\,0.03,1.5\times 10^{-3}\}$.}
\label{profbi4}
\end{figure}

We consider three different values for the resistivity, corresponding to a Lundquist number at the Jeans scale defined by $S|_{q=q_J}=\{10,\,0.3,0.07\}$. In Figure \ref{profbi4}, we show the equal density contours where $\delta= 0.99 \delta_c$.
As a result, we show how the bigger the resistivity is the lower is the ``squeeze'' of the equal density contours along the orthogonal direction. In this respect, we conclude noting how the resistivity  reduces the anisotropy in the density perturbations due to the presence of the magnetic field.

\section{Discussion and Conclusions}\label{Sec:fi}
We have considered the stability of a static, homogeneous infinite plasma. The perturbation equations and the stability problem has been studied for this system in some details and the perturbation frequencies have been classified, pointing out different regions which are marked by the values of the resistivity and Alfv\`en velocity. Two distinct regimes have been addressed: the ideal case and a resistive MHD scheme.

In the ideal MHD picture, it has been outline how, if the propagation vector lies on the plane orthogonal to the background magnetic field, the stability is affected by the presence of the magnetic pressure, and, since we obtain $v_s^2\to v_s^2 + v_a^2$, the collapsing characteristic scale is greater than the standard Jeans one. On the other hand, out of this plane, the standard Jeans criterion remains valid. The anisotropy in the density-perturbation profile has been shown as a consequence of the presence of the magnetic field: as time goes by, the equal density contours squeeze along the direction orthogonal to the background magnetic field.

For the non-vanishing resistivity regime, the situation can be summarized as follows. The standard Jeans criterion governs the stability of all perturbations, irregardless of their direction. In fact, for small resistivity, the magnetic field lines are frozen into the fluid and the field--perpendicular motion will therefore drag the lines with it, and increase the magnetic pressure, while the field--parallel compression does not increase the magnetic field strength. However, in the limit of very small but non-zero resistivity, it is found that the speed of the perturbation growth is strongly direction-dependent. In particular, perturbations in the orthogonal plane grow much slowly with respect to those parallel to the magnetic field. This again leads to an anisotropy in the gravitational collapse similar to that observed in the zero-resistivity case. On the other hand, the difference in the evolution of the density perturbation profile along the orthogonal direction, as shown in the zero and small resistivity cases, disappears as the resistivity increases, suggesting a balance effect between the magnetic field and the dissipative term driven by resistivity.

In this work, the resistive MHD system has been addressed as a simplified (one-fluid viscosity--free MHD) model describing the astrophysical plasmas. Indeed, one should complete this framework providing a generalization towards a more realistic models, \emph{e.g.}, visco--resistive MHD. Such scheme, allows to include the dissipation processes driven by viscosity and constitutes the basis of accretion disk models, since the (turbulent) visco-resistive plasma configurations can be regard as one of principal element for the momentum transport across the disk \cite{Bisno01}. Moreover, a two--fluids description (with ions and electrons) or multi-fluids models with charged and neutral particles should account for other effects like the ambipolar diffusion, important for many astrophysical processes that involve lightly ionized gas.

\section{Acknowledgment}
This work has been developed in the framework of the CGW Collaboration (www.cgwcollaboration.it). DP gratefully acknowledges financial support from the Angelo Della Riccia Foundation. NC would like to acknowledge the Centre de Physique Th\'eorique, Universit\'e de la Mediterran\'ee Aix-Marseille 2 and the financial support from ``Sapienza'' University of Rome. ML acknowledges financial support from a joint Accademia dei Lincei / Royal Society fellowship for Astronomy.

\addcontentsline{toc}{chapter}{Bibliography}

\newcommand{\nar}{New Astronomical Reviews}

\bibliography{Jeans_R1}
\bibliographystyle{model1a-num-names}

\end{document}